\documentclass[baaa]{baaa}

\usepackage[pdftex]{hyperref}
\usepackage{subfigure}
\usepackage{natbib}
\usepackage{helvet,soul}
\usepackage[font=small]{caption}

\begin{document}

%%%%%%%%%%%%%%%%%%%%%%%%%%%%%%%%%%%%%%%%%%%%%%%%%%%%%%%%%%%%%%%%%%%%%%%%%%%%%%
%                                                                            %
% Datos de la publicación, no deben ser cambiados.                           %
%                                                                            %
% Journal data, please do not change them.                                   %
%                                                                            %
%%%%%%%%%%%%%%%%%%%%%%%%%%%%%%%%%%%%%%%%%%%%%%%%%%%%%%%%%%%%%%%%%%%%%%%%%%%%%%

\journalvol{61A}
\journalyear{2019}
\journaleditors{R. Gamen, N. Padilla, C. Parisi, F. Iglesias \& M. Sgr\'o}

%%%%%%%%%%%%%%%%%%%%%%%%%%%%%%%%%%%%%%%%%%%%%%%%%%%%%%%%%%%%%%%%%%%%%%%%%%%%%%
%                                                                            %
%  Seleccione el idioma de su contribución: Recuerde que todos los           %
%  componentes del documento (titulo, texto, figuras, tablas, etc.)          %
%  deben estar en el mismo idioma.                                           %
%                                                                            %
%  Select the languague of your contribution: Please remember that all       %
%  document parts (title, text, figures, tables, etc.) must be in the        %
%  same languaje.                                                            %
%                                                                            %
%  0: Castellano / Spanish                                                   %
%  1: Inglés / English                                                       %
%                                                                            %
%%%%%%%%%%%%%%%%%%%%%%%%%%%%%%%%%%%%%%%%%%%%%%%%%%%%%%%%%%%%%%%%%%%%%%%%%%%%%%

\contriblanguage{1}

%%%%%%%%%%%%%%%%%%%%%%%%%%%%%%%%%%%%%%%%%%%%%%%%%%%%%%%%%%%%%%%%%%%%%%%%%%%%%%
%                                                                            %
%  Seleccione el tipo de contribución solicitada:                            %
%                                                                            %
%  Select the requested contribution type:                                   %
%                                                                            %
%  1: Presentación mural / Poster                                            %
%  2: Presentación oral / Oral contribution                                  %
%  3: Informe invitado / Invited report                                      %
%  4: Mesa redonda / Round table                                             %
%  5: Presentación Premio Varsavsky / Varsavsky Prize contribution           %
%  6: Presentación Premio Sahade / Sahade Prize contribution                 %
%  7: Presentación Premio Sérsic / Sérsic Prize contribution                 %
%                                                                            %
%%%%%%%%%%%%%%%%%%%%%%%%%%%%%%%%%%%%%%%%%%%%%%%%%%%%%%%%%%%%%%%%%%%%%%%%%%%%%%

\contribtype{2}

\thematicarea{1}

\title{Impact of radiation backgrounds on the formation of massive black holes}
\subtitle{}

%%%%%%%%%%%%%%%%%%%%%%%%%%%%%%%%%%%%%%%%%%%%%%%%%%%%%%%%%%%%%%%%%%%%%%%%%%%%%%
%                                                                            %
%  Agregue un título corto para el encabezado de las páginas pares.          %
%                                                                            %
%  Add a short title to appear in the header of even pages.                  %
%                                                                            %
%%%%%%%%%%%%%%%%%%%%%%%%%%%%%%%%%%%%%%%%%%%%%%%%%%%%%%%%%%%%%%%%%%%%%%%%%%%%%%

\titlerunning{UV backgrounds effect on SMBHs formation}

%%%%%%%%%%%%%%%%%%%%%%%%%%%%%%%%%%%%%%%%%%%%%%%%%%%%%%%%%%%%%%%%%%%%%%%%%%%%%%
%                                                                            %
%  Lista de autores. Los nombres de los autores deben estar separados por    %
%  comas, y deben tener el formato A.E. Autor (iniciales apellido(s);   sin  %
%  coma entre apellido e iniciales ni espacios entre las iniciales).         %
%                                                                            %
%  Author list. Authors' names must be separated by commas, and stick to     %
%  the format A.E. Author (initials Family name -neither commas between      %
%  name and the initials nor blanks between the initials).                   %
%                                                                            %
%%%%%%%%%%%%%%%%%%%%%%%%%%%%%%%%%%%%%%%%%%%%%%%%%%%%%%%%%%%%%%%%%%%%%%%%%%%%%%

\author{V.B. Díaz\inst{1}, D.R.G. Schleicher\inst{1}, S. Bovino\inst{1}, FFibla\inst{1}, R. Riaz\inst{1}, S. Vanaverbeke\inst{2} \& C. Olave\inst{1}}
\authorrunning{Díaz et al.}

%%%%%%%%%%%%%%%%%%%%%%%%%%%%%%%%%%%%%%%%%%%%%%%%%%%%%%%%%%%%%%%%%%%%%%%%%%%%%%
%                                                                            %
% Por favor provea una dirección de e-mail de contacto para los lectores.    %
%                                                                            %
% Please provide a contact e-mail address for the readers.                   %
%                                                                            %
%%%%%%%%%%%%%%%%%%%%%%%%%%%%%%%%%%%%%%%%%%%%%%%%%%%%%%%%%%%%%%%%%%%%%%%%%%%%%%

\contact{vdiazd@udec.cl}

\institute{
Departamento de Astronom\'ia, Facultad de Ciencias F\'isicas y Matem\'aticas, Universidad de Concepci\'on, Av.Esteban Iturra s/n Barrio Universitario, Casilla 160-C, Concepci\'on, Chile \and
Centre for mathematical Plasma-Astrophysics, Department of Mathematics, KU Leuven, Celestijnenlaan 200B, 3001 Heverlee, Belgium
}

%%%%%%%%%%%%%%%%%%%%%%%%%%%%%%%%%%%%%%%%%%%%%%%%%%%%%%%%%%%%%%%%%%%%%%%%%%%%%%
%                                                                            %
%  El resumen y el abstract son ambos obligatorios, independientemente del   %
%  lenguaje elegido.                                                         %
%                                                                            %
%  The Resumen and the abstract are both mandatory, regardless of the chosen %
%  language.                                                                 %
%                                                                            %
%%%%%%%%%%%%%%%%%%%%%%%%%%%%%%%%%%%%%%%%%%%%%%%%%%%%%%%%%%%%%%%%%%%%%%%%%%%%%%

\resumen{
La existencia de agujeros negros supermasivos de mil millones de masas solares a muy alto corrimiento al rojo nos ha motivado a estudiar como estos objetos tan masivos se forman durante los primeros miles de millones de años después del Big Bang. El modelo mas prometedor que se ha propuesto es el colapso directo de nubes de gas protogalácticas. Este escenario requiere altas tasas de acreción para crear rápidamente objetos masivos y la inhibición del enfriamiento que causa $\mathrm H_2$, el cuál es importante en el proceso de fragmentación.  Estudios recientes mostraron que, si usamos un fondo radiativo fuerte, el hidrógeno molecular se destruye, favoreciendo las altas tasas de acreción y por lo tanto formando objetos de muy alta masa. En este trabajo estudiamos el impacto de campos de radiación UV en una nube de gas primordial usando el código GRADSPH-KROME para investigar el proceso de fragmentación en escalas de unidades astronómicas y por lo tanto la formación de los primeros agujeros negros supermasivos. Encontramos que para suprimir la formación de $\mathrm H_2$ es necesario un valor de $J_{21}$ muy alto, por lo que los agujeros negros de colapso directo no podrían explicar la formación de los primeros agujeros negros supermasivos.}

\abstract{
The presence of supermassive black holes (SMBHs) of a few billion solar masses at very high redshift has motivated us to study how these massive objects formed during the first billion years after the Big Bang. The most promising model that has been proposed to explain this is the direct collapse of protogalactic gas clouds. In this scenario, very high accretion rates are needed to form  massive objects early on and the suppression of $\mathrm H_2$ cooling is important in regulating the fragmentation. Recent studies have shown that if we use a strong radiation background, the hydrogen molecules are destroyed, favoring the high accretion rates and therefore producing objects of very high mass. In this work we study the impact of UV radiation fields in a primordial gas cloud using the recently coupled code GRADSPH-KROME for the modeling of gravitational collapse including primordial chemistry to explore the fragmentation in AU scales and hence the formation of first SMBHs. We found that to suppress the formation of $\mathrm H_2$ a very high value of $J_{21}$ is required, because of that we conclude that the direct collapse black holes (DCBHs) are very unlikely to be an explanation for the formation of the first SMBHs. 
}

%%%%%%%%%%%%%%%%%%%%%%%%%%%%%%%%%%%%%%%%%%%%%%%%%%%%%%%%%%%%%%%%%%%%%%%%%%%%%%
%                                                                            %
%  Seleccione las palabras clave que describen su contribución. Las mismas   %
%  son obligatorias, y deben tomarse de la lista de la American Astronomical %
%  Society (AAS), que se encuentra en la página web indicada abajo.          %
%                                                                            %
%  Select the keywords that describe your contribution. They are mandatory,  %
%  and must be taken from the list of the American Astronomical Society      %
%  (AAS), which is available at the webpage quoted below.                    %
%                                                                            %
%  https://aas.org/authors/astronomical-subject-keywords-update-august-2013  %
%                                                                            %
%%%%%%%%%%%%%%%%%%%%%%%%%%%%%%%%%%%%%%%%%%%%%%%%%%%%%%%%%%%%%%%%%%%%%%%%%%%%%%


\maketitle

\section{Introduction}
\label{S_intro}
More than $100$ supermassive black holes (SMBHs) with masses of about $10^9$~M$_\odot$ at very high redshift ($z\geq6$) have been discovered in the last years through several surveys \citep{gallerani,domi2018}. The highest-redshift quasar observed is at $z=7.54$ with a mass of $8\times10^8$~M$_\odot$ \citep{banados} and another one at $z=7.085$ with a mass of $2\times10^9$~M$_\odot$ \citep{morlock}. The formation of the first structures is not yet understood, and the formation of the first supermassive black holes is still an open question in cosmology. Among the models that have been proposed to explain the formation of SMBHs in the early universe, the direct collapse of protogalactic gas clouds \citep{loeb,brom,sglosman} is the most promising scenario as it provides the most massive black holes seeds ($M\sim 10^5$~M$_\odot$) which can then grow at relatively moderate accretion rates to form SMBHs.

The formation of direct collapse black holes (DCBHs) requires an efficient accretion rate of gas to the central object ($\dot{M} \approx 1$ M$_{\odot}\;$yr$^{-1}$) and the suppression of fragmentation of the cloud. These conditions can be achieved if the gas collapses isothermally at a temperature of $T \approx 10^4$ K \citep{omukai}. Such a collapse is possible if the gas has zero metallicity and the main cooling mechanism in the early universe ($\mathrm H_2$ cooling) is suppressed due to an intense radiation background \citep{brom, visbal}.

In order to study the cooling process of the gas cloud, we need to include the chemical reactions involving the formation (via gas-phase reactions) %and destruction 
of $\mathrm H_2$:
\begin{align}
\label{eq:ecuacion1}
 \mathrm H + \mathrm e^- &\to \mathrm H^- + \gamma\\
 \mathrm H+\mathrm H^-&\to \mathrm H_2 +\mathrm e^-
\end{align}

Once the first generation of stars (Pop III) are formed, they will irradiate the intergalactic medium (IGM) with a UV flux
and pollute it with metals through supernova explosions leading to the formation of the second generation of stars (Pop II).  The UV flux produced by these stellar populations can destroy $\mathrm H_2$ through the Solomon process (Eq.~\ref{eq:ecuacion3}) and photo-detach electrons from $\mathrm H^-$ (Eq.~\ref{eq:ecuacion4}). 
\begin{align}
\label{eq:ecuacion3}
\mathrm H_2+ \gamma_{LW}&\to \mathrm H+\mathrm H\\
\label{eq:ecuacion4}
\mathrm H^- + \gamma_{0.76} &\to \mathrm H + \mathrm e^-
\end{align}

Thus, massive primordial haloes of  $10^7 - 10^8 $~M$_{\odot}$ which formed in the early universe and irradiated by nearby star-forming regions of Pop II and Pop III at $z=15-20$ are the most plausible cradles for DCBH formation. The available flux from star-forming regions in measured in units of $J_{21}$, which $J_{21}=1$ corresponding to a flux of $10^{-21}$~erg~cm$^{-2}$~s$^{-1}$~Hz$^{-1}$~sr$^{-1}$ at the Lyman limit. We assume for simplicity that the shape of the spectrum corresponds to a blackbody.

In a recent study, \citet{latif4} performed 3D cosmological simulations to determine the critical UV flux $J^{\rm crit}_{21}$
above which $\mathrm H_2$ cooling is suppressed in protogalactic gas clouds of $10^7-10^8$~M$_{\odot}$ including the impact of X-ray ionization and realistic Pop II spectra. They found that $J^{\rm crit}_{21}$ for realistic Pop II spectra is a few times $10^4$ and weakly depends on the adopted radiation spectra in the range between $T_{\rm rad} = 2\times 10^4-10^5$~K and the impact of X-ray ionization is negligible. These results suggest that DCBHs could be rarer than previously thought.

\section{Computational Methods}
In this work we performed our simulations with the coupling of the Smoothed Particle Hydrodynamics (SPH) code GRADSPH\footnote{\url{http://www.swmath.org/software/1046}} \citep{gradsph} with the chemistry package KROME\footnote{\url{http://www.kromepackage.org}} \citep{krome}. This combined code called GRADSPH-KROME allows us to include the chemistry and cooling in hydrodynamical simulations of the star forming gas. The code was previously employed by \citet{rafeel1} to explore the fragmentation process for the formation of binary systems. Here we explore the chemical conditions in the presence of different UV fluxes, to determine the flux that is required for an atomic collapse.
\subsection{GRADSPH}
SPH is a mesh-free Lagrangian method used for simulating the dynamics of continuous media such us fluid flows. It works by dividing the fluid into a set of discrete and spherically symmetric particles. Each particle has associated with it a mass $m_i$, a velocity vector $\vec{v}_i$, and values of thermodynamic variables which describe the state of a fluid, such as the pressure $P_i$, the density $\rho_i$, and the specific internal energy $u_i$. Also, these particles are associated with a spatial scale known as the smoothing length $h_i$, over which their properties are smoothed by a kernel or weighing function $W$. So, any property can be obtained by summing the relevant properties of all the particles which lie within the range of the kernel. Hence, the density $\rho_i$ at the position $\mathbf{r}_i$ of each particle with mass $m_i$ is determined by
\begin{align}
\rho_i=\sum_jm_jW(\mathbf{r}_i-\mathbf{r}_j,h_i).
\end{align}

GRADSPH uses the standard M4-kernel or cubic spline kernel  with a compact support that contains particles within a smoothing sphere of size $2h_i$ \citep{price1}. This smoothing length is determined by $h_i=\eta \left(m_i/\rho_i\right)^{1/3}$
%\begin{align}
%\label{eq:ecuacion6}
%h_i=\eta \left(\frac{m_i}{\rho_i}\right)^{1/3},
%\end{align}
where $\eta$ is a dimensionless parameter which determines the size of the smoothing length of the SPH particle given its mass and density and is determined by the following equation:
 \begin{align}
 \label{eq:ecuacion7}
 \eta=\frac{1}{8}\left(\frac{3N_{opt}}{4\pi}\right)^{1/3},
 \end{align}
 where $N_{opt}$ is the number of neighbors inside the smoothing sphere, which can be between 50 to 100. In this work we use $N_{opt}=50$ for the 3D-simulations. Also, it is important to know 
 that the mass contained within the smoothing sphere of each particle should be held constant.
 
%  \begin{align}
%\frac{d\mathbf{v}_i}{dt}=&-\sum_{j=1}^{N}m_j\left(\frac{P_i}{\rho_i^2\Omega_i}\nabla_i W(r_{ij},h_i)\right\nonumber\\
%&+\left\frac{P_j}{\rho_j^2\Omega_j}\nabla_i W(r_{ij},h_j) \right),
 %\label{eq:ecuacion8}
 %\end{align}
%in which the coefficients $\Omega_i$ are defined by
 %\begin{align}
 %\label{eq:ecuacion9}
%\Omega_i=1-\frac{\partial h_i}{\partial \rho_i}\sum_{j=1}^{N}m_j\frac{\partial W(r_{ij},h_j)}{\partial h_j}.
% \end{align}
 
Also, GRADSPH implemented a second-order PEC (predict-evaluate-correct) scheme combined with  an individual particle time stepping method to solve the system of ordinary differential equations
that updates the positions and velocities of the particles. The work presented in \citet{rafeel2} and \citet{rafeel3} are examples of simulations with GRADSPH.
%\subsection{Sink particle treatment}
%Star and black hole formation is such a complex process, because of this, accurate numerical tools are needed to quantitatively examine the mass distribution and accretion of fragments in collapsing and turbulent gas clouds. Thus, in our simulations, protostars are represented by sink particles that follows the creation critera postulated by \citet{sink}. These are:
%\begin{itemize}
  %  \item The density at the location of the particle should exceed an user-defined threshold, which is set to $\rho_{SINK}=10^{-15}$~g/cm$^3$.
   % \item the sink at the moment of creation, should not overlap another sink.
    %\item the gravitational potential should be lower that of all it neighbors.
    %\item the density should satisfy the Hill condition
    %\begin{eqnarray}
%\rho_i>\rho_{HILL}\equiv \frac{3X_{HILL}(-\Delta\mathbf{r}_{is'}\cdot\Delta\mathbf{a}_{is'})}{4\pi G |\Delta\mathbf{r}_{is'}|^2}
%\end{eqnarray}
%where $X_{hill}$ is a user-defined parameter with default value of 4, $\Delta\mathbf{r}_{is'}$ and $\Delta\mathbf{a}_{is'}$ are the position and acceleration of particle i relative to the pre-existing sink.
%\end{itemize}
\subsection{Chemistry, cooling and UV background}
The KROME package allows us to model chemical network in numerical simulations. In this work, we prepare a chemical network based on the network react\_xrays provided by KROME with the chemical reactions presented in \citet{glover1} and \citet{glover2}, giving a total of 35 chemical reactions with 9 different chemical species ($\mathrm e^-$, $\mathrm H^-$, H, $\mathrm H^+$, He, $\mathrm He^+$, $\mathrm He^{++}$, $\mathrm H_2$, $\mathrm H_2^+$). The initial mass fraction of these chemical species are: $f_{\mathrm H}=0.75$, $f_{\mathrm H_e}=0.24899$, $f_{\mathrm H_2}=10^{-3}$, $f_{\mathrm H}=8.2\times10^{-4}$, $f_{\mathrm e^-}=4.4\times10^{-8}$, the other species are set zero. To solve the rate equations, KROME has a main module that calls the high-order solver DLSODES. To include the presence of a UV background in GRADSPH-KROME we add new files generated by KROME to the network that includes the chemical reactions for photodissociation and photodetachment of $\mathrm H_2$ due to a UV background and also we set the function krome\_set\_user\_J21(J21) for the values of $J_{21}$.
\subsection{Setup}
%\begin{table}[!t]
%\centering
%\caption{Initial conditions for both  atomic cooling and UV background simulations.}
%\begin{tabular}{lc}
%\hline\hline\noalign{\smallskip}
%Parameter & Value\\
%\hline\noalign{\smallskip}
%Number of particles ($N$) & $\sim500\;000$\\
%Mass of the cloud ($M_{cloud}$)& $6.4\times10^6$~M$_{\odot}$\\
%Radius of the cloud ($R_{cloud}$) & $80.4$~pc\\
%Density of the cloud ($\rho_{cloud}$) & $2.0\times10^{-22}$~g/cm$^3$\\
%Temperature ($T$) & $10^4$~K\\
%Mach number ($\mathcal{M}$) & $1$\\
%Angular velocity ($\omega$) & $2.3\times10^{-15}$~rad/s\\
%\hline
%\end{tabular}
%\label{tab:table1}
%\end{table}
Our spherical primordial gas cloud is modeled as a distribution of $507\;443$ SPH particles with an initial temperature of $10^4$~K. This cloud has a total mass of $M_{cloud}=6.4\times10^6$~M$_{\odot}$, a radius of $R_{cloud}=80.4$~pc and therefore an initial density of $\rho_{cloud}=2.0\times10^{-22}$~g$\;$cm$^{-3}$ also, the gas is in solid body rotation with  an angular velocity of $\omega=2.3\times10^{-15}$~rad$\;$s$^{-1}$ and is turbulent with a Mach number $\mathcal{M}=1.0$.
\section{Results}
\label{Results}
\begin{figure}[!t]
  \centering
  \includegraphics[width=0.4679\textwidth]{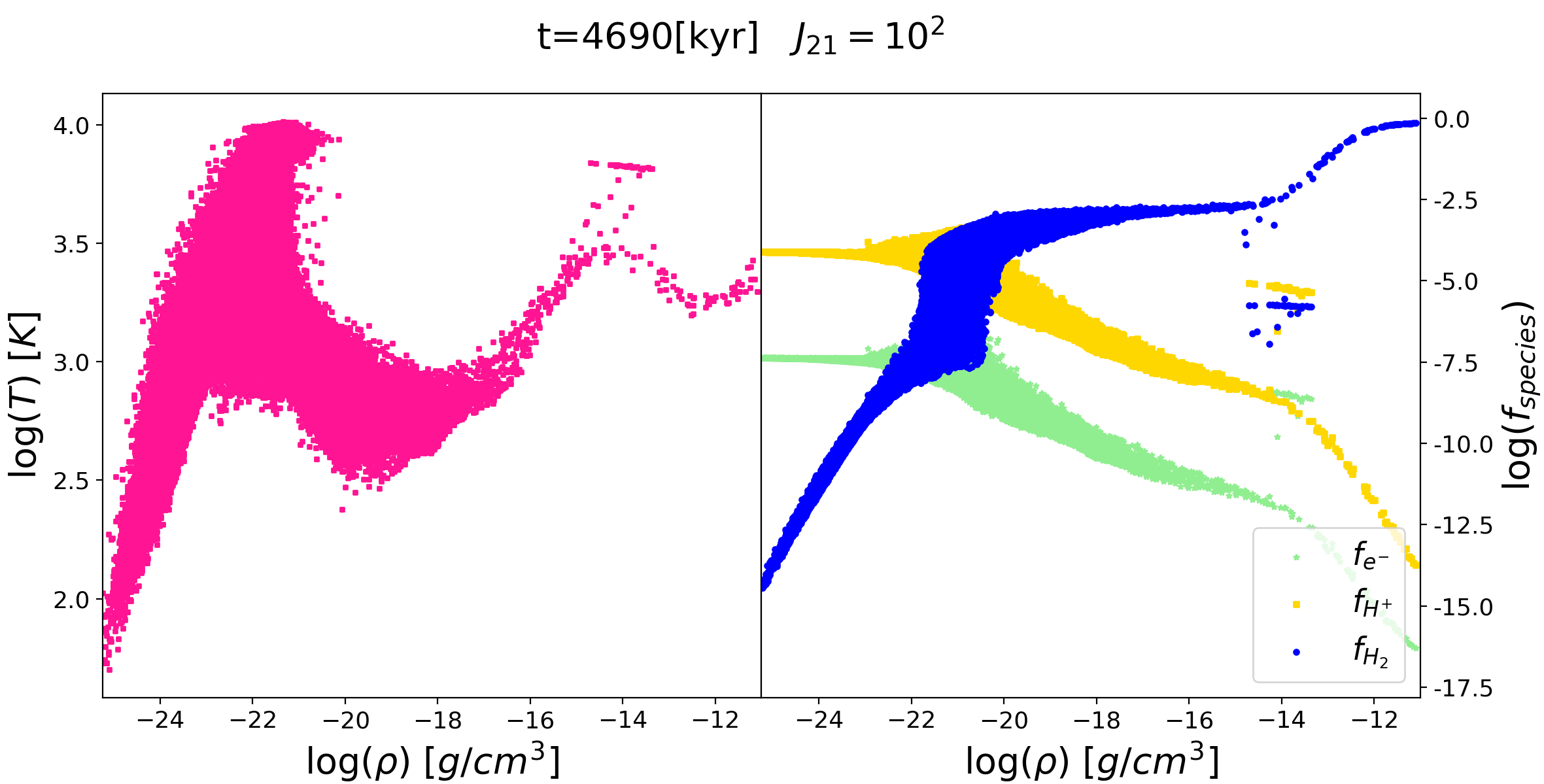}
  \includegraphics[width=0.4679\textwidth]{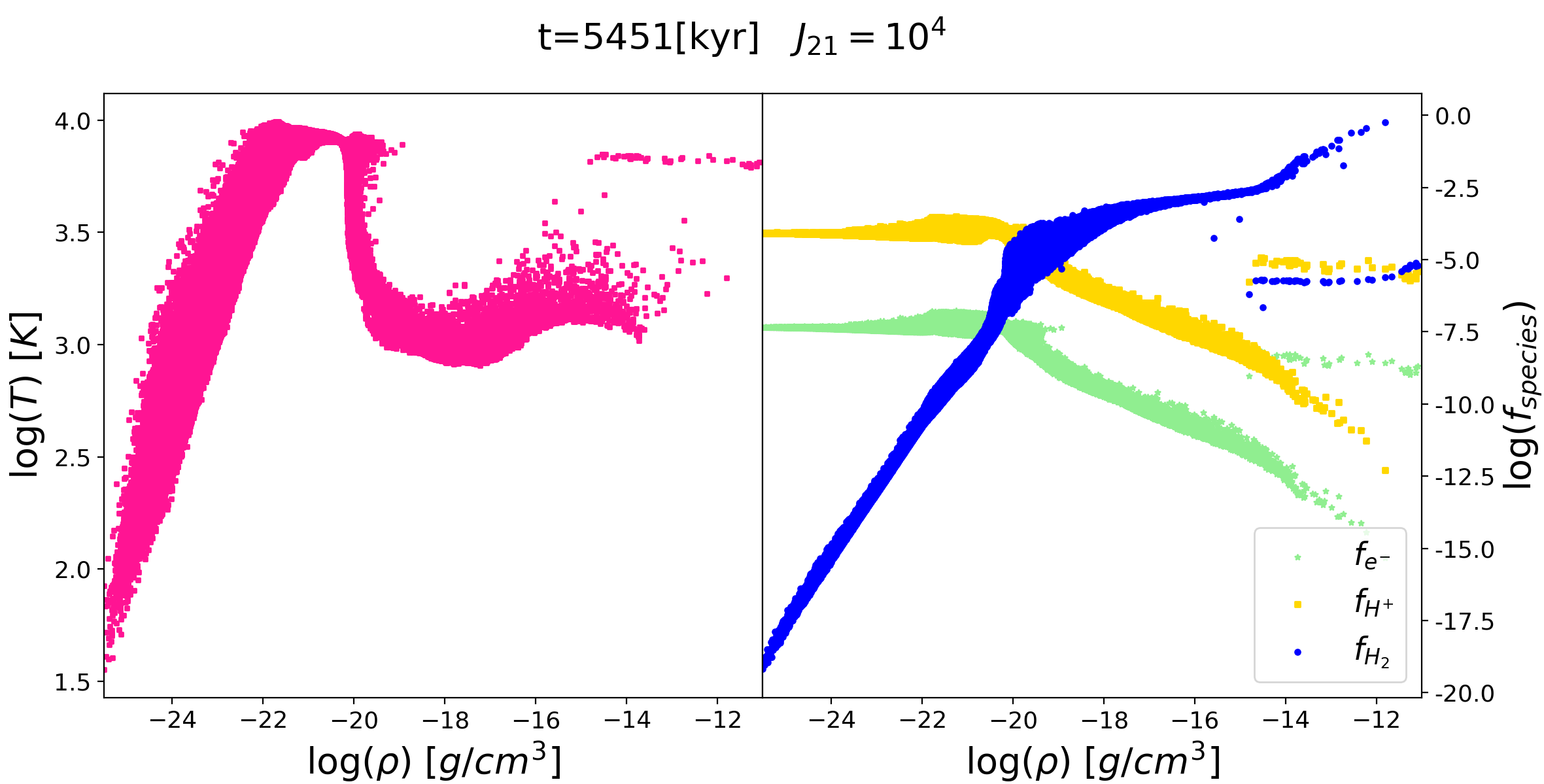}
  \includegraphics[width=0.4679\textwidth]{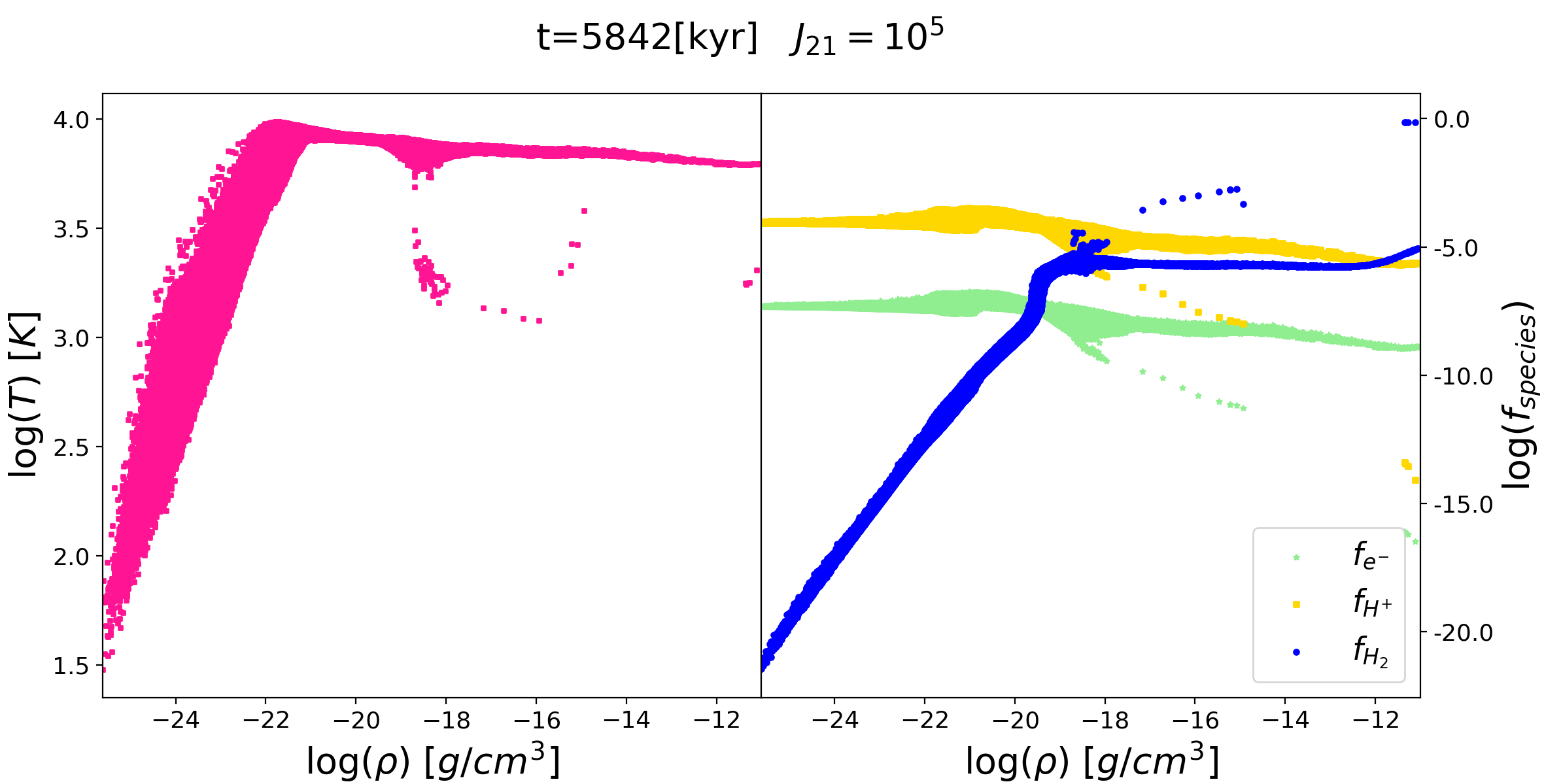}
  \caption{Thermal and species profiles ($\mathrm H_2$ in blue dots, $\mathrm H^+$ in yellow squares and $\mathrm e^{-}$ in green stars) for different strengths of UV flux. Upper panel is for $J_{21}=10^2$ middle panel for $J_{21}=10^4$ and bottom panel for $J_{21}=10^5$.}
  \label{F_letras}
\end{figure}

Fig. \ref{F_letras} shows the thermal and species profiles for different strengths of the UV flux. We can see that for the weaker value of $J_{21}$, the cooling due to $\mathrm H_2$ becomes effective, in which the gas is initialized with a temperature of $10^4$~K and then cools down to about $10^3$~K. For $J_{21}=10^4$  we can see that the $\mathrm H_2$ formation remains suppressed until a density of $10^{-20}$~g$\;$cm$^{-3}$ which illustrates the presence of two gas phases at the same density similar to Fig. 3 from \citet{latif}. For the stronger value of $J_{21}$ we see that the gas is in an atomic state in which $\mathrm H_2$ remains suppressed due to the high radiation and it remains in a hot phase, according to theoretical expectations. Nevertheless, the value of $J^{\rm crit}_{21}$ that we found is in the range of $10^4-10^5$ which is very high. Also, the right panel of Fig.\ref{F_letras} shows the evolution of $\mathrm H_2$,  $\mathrm H^+$ and $\mathrm e^{-}$. As they act as catalysts we see that for the weaker values of $J_{21}$, the amount of $\mathrm e^{-}$ and $\mathrm H^+$ is the same after the reaction and recombination have happened and their number densities are depleted due to the formation of $\mathrm H_2$. For $J_{21}=10^5$ the $\mathrm H^+$ and $\mathrm e^{-}$ number densities increase and become constant because the formation of $\mathrm H_2$ remains inhibited.

%%%%%%%%%%%%%%%%%%%%%%%%%%%%%%%%%%%%%%%%%%%%%%%%%%%%%%%%%%%%%%%%%%%%%%%%%%%%%%
%                                                                            %
% Para figuras de dos columnas use \begin{figure*} ... \end{figure*}         %
%                                                                            %
%%%%%%%%%%%%%%%%%%%%%%%%%%%%%%%%%%%%%%%%%%%%%%%%%%%%%%%%%%%%%%%%%%%%%%%%%%%%%%

\section{Conclusions and Outlook}
\label{Conclusions and Outlock}
We found that the required value of $J_{21}$ to keep the gas atomic is between $10^4-10^5$, because of that we conclude that black hole formation via direct collapse is efficiently suppressed, and cannot explain the observed quasars at high redshift. This conclusion is also found in \citet{Dijkstra}, in which they found that the value of the comoving number density of putative DCBHs formation sites increases with the cosmic time but to obtain the necessary high values of $J_{21}$ for these sites we have to consider very nearby star-forming galaxies including galactic winds that will produced metal enrichment, which will suppress the predicted putative DCBHs formation sites and they do not expect this formation channel to be sufficient enough, however, an alternative pathway to form SMBHs is the collision in primordial star clusters \citep{reinoso,tjarda}.

%In the future, we plan to study in more detail the fragmentation behaivour of intermediate values of J_{21}.

\begin{acknowledgement}
This research was partially supported by the supercomputing infrastructure of the NLHPC (ECM-02), also the authors acknowledge the Kultrun Astronomy Hybrid Cluster (projects Conicyt Programa de Astronomia FondoQuimal QUIMAL170001, Conicyt PIA ACT172033, and Fondecyt Iniciacion 11170268) for providing HPC resources that have contributed to the research results reported in this paper. VBD thanks to Conicyt for financial support on her master studies (CONICYT-PFCHA/Mag\'isterNacional/2017-22171293). DRGS, SB, FF, CO, RR and VBD thank for funding via CONICYT PIA ACT172033. FF and VBD thank for funding through Fondecyt regular (project code 1161247). RR, CO, FF and DRGS thank for funding through the 'Concurso Proyectos Internacionales de Investigaci\'on, Convocatoria 2015' (project code PII20150171). DRGS and SB acknowledge funding through CONICYT project Basal AFB-170002.
%the Geryon/Geryon2 cluster housed at the Centro de Astro-Ingenieria UC was used for (part) the calculations performed in this paper. The BASAL PFB-06 CATA, Anillo ACT-86, FONDEQUIP AIC-57, and QUIMAL 130008 provided funding for several improvements to the Geryon/Geryon2 cluster. The authors acknowledge the Kultrun Astronomy Hybrid Cluster (projects Conicyt Programa de Astronomia FondoQuimal QUIMAL170001, Conicyt PIA ACT172033, and Fondecyt Iniciacion 11170268) for providing HPC resources that have contributed to the research results reported in this paper. VD thanks to Conicyt for financial support on her master studies (CONICYT-PFECHA/Mag\'isterNacional/2017-22171293)
%Agradecemos a todos los miembros de los Comités Organizadores Local y 
%Científico por su activa
%participación que permitió llevar a cabo una exitosa reunión. 
\end{acknowledgement}

%%%%%%%%%%%%%%%%%%%%%%%%%%%%%%%%%%%%%%%%%%%%%%%%%%%%%%%%%%%%%%%%%%%%%%%%%%%%%%
%                                                                            %
%  Por favor no modifique las líneas de la bibliografía, salvo el nombre     %
%  el archivo de Bibtex con la lista de citas (sin la extensión .BIB)        %
%                                                                            %
%  Please do not modify the following lines, except the name of the Bibtex   %
%  file (whithout the .BIB extension)                                        %
%                                                                            %
%%%%%%%%%%%%%%%%%%%%%%%%%%%%%%%%%%%%%%%%%%%%%%%%%%%%%%%%%%%%%%%%%%%%%%%%%%%%%% 

\bibliographystyle{baaa}
\small
\bibliography{biblio}
 
\end{document}